# Verification, Validation, and Uncertainty Quantification (VVUQ) of Physiologically Based Pharmacokinetic Models for Theranostic Digital Twins: Towards Reliable Model-Informed Treatment Planning for Radiopharmaceutical Therapies

Nouran R. R. Zaid[1], PhD, Deni Hardiansyah[2], PhD, Tahir Yusufaly [1], PhD, Arman Rahmim[3,4,5], PhD

[1]Department of Radiology and Radiological Science, Johns Hopkins University School of Medicine, Baltimore, MD, USA; [2]Medical Physics and Biophysics, Physics Department, Faculty of Mathematics and Natural Sciences, Universitas Indonesia, Depok, Indonesia; [3]Department of Integrative Oncology, BC Cancer Research Institute, Vancouver, BC, Canada. [4]Departments of Radiology and Physics, University of British Columbia, Vancouver, BC, Canada, [5]School of Biomedical Engineering, University of British Columbia, Vancouver, BC, Canada

**First and corresponding Author:** Nouran R. R. Zaid, 1550 Orleans St, Baltimore, MD 21287, *410-614-0254*, nzaid1@jh.edu

**Co-authors:** denihardiansyah@sci.ui.ac.id, tyusufa2@jhmi.edu, arman.rahmim@ubc.ca

## Key Points

- Digital twins (DTs) have the potential to advance personalized radiopharmaceutical therapies (RPTs) beyond current one-size-fits-all paradigms.
- DTs, especially those informed by theranostics (leading to theranostic DTs; TDTs), are founded on bi-directional interactions between personalized *in silico* models, such as physiologically based pharmacokinetic (PBPK) models, and patient-specific real-time data inputs, including biokinetic data from theranostic imaging, to enable individualized dosimetry.

- Rigorous verification, validation, and uncertainty quantification (VVUQ) is required to ensure model reliability for clinical decision-making, which we discuss in this work in the context of PBPK models.
- The dynamic real-time updating of TDTs introduces additional variabilities, mandating data-responsive VVUQ practices to ensure reliability and manage uncertainty during the course of treatment.
- Regulatory alignment and society-led standards will be vital for clinical adoption of TDTs in RPTs.

**Synopsis**

Physiologically based pharmacokinetic (PBPK) models provide a mechanistic framework for simulating radiopharmaceutical kinetics and estimating patient-specific absorbed doses (ADs). PBPK models incorporate prior knowledge of patient physiology and drug-specific properties, which can enhance the model's predictive performance. PBPK models can ultimately be used to predict treatment response and thereby enable theranostic digital twins (TDTs) for personalized treatment planning in radiopharmaceutical therapies (RPTs). To achieve this potential of precision RPT, however, the reliability of the underlying modeling, including the PBPK-based dosimetry, must be established through rigorous verification, validation, and uncertainty quantification (VVUQ). This review outlines the role of VVUQ in ensuring the credibility and clinical applicability of PBPK models in radiotheranostics. Key methodologies for PBPK model VVUQ are discussed, including goodness-of-fit (GOF) assessment, prediction evaluation, and uncertainty propagation.

## 1. **Introduction**

Radiotheranostics is a powerful paradigm for personalized cancer treatment in radiopharmaceutical therapies (RPTs) (1). It relies on agents that bind to specific biological targets, thereby enabling simultaneous diagnostic imaging and therapeutic intervention (2). The clinical impact of radiotheranostic agents is highlighted by their expanding role in cancer treatment, as reflected in the Food and Drug Administration (FDA) approval of [$^{177}$Lu]Lu-PSMA-617 for early treatment of prostate cancer (3, 4).

Despite the potential of quantitative targeted molecular imaging to guide and optimize personalized treatment planning in RPT, current RPT regimens remain largely standardized, with fixed administered activities and treatment intervals (5). This one-size-fits-all approach overlooks interindividual variability in numerous factors, such as pharmacokinetics, tumor burden, and radiobiological sensitivity, leading to considerable differences in therapeutic efficacy and toxicity across populations (6). Computational nuclear oncology (CNO) is a recently proposed model-informed treatment planning paradigm for RPTs using theranostic digital twins (TDTs). Among the various TDT implementations in CNO, physiologically based pharmacokinetic (PBPK) models offer distinct advantages over conventional sum-of-exponential or compartmental models by incorporating prior knowledge of patient physiology and drug characteristics, thereby enabling more accurate estimation of pharmacokinetic (PK) and radiobiological parameters. The PBPK models can estimate ADs and clinical responses to both target and normal tissues simultaneously (7-9), providing a systematic strategy that accounts for interpatient variability and optimizing treatment outcomes. Ultimately, such a modeling promises to facilitate the transition to model-informed precision RPTs via personalized, dosimetry-driven decision-making (Mathematical and Computational Nuclear Oncology: Toward Optimized Radiopharmaceutical Therapy via Digital Twins, Marc Ryhiner et al., PET Clinics, 2026).

The application and prospective evolution of TDTs into RPTs critically depend on the quality and seamless integration of personalized data into *in silico* models, enabling a bi-directional exchange of information in which patient-specific inputs inform the model, and the model outputs, in turn, guide personalized treatment planning. However, the lack of standardized approaches for model structure, parameter estimation, and data integration introduces variability and uncertainty in model predictions. This variability, compounded by the commonly available sparse clinical datasets, undermines the reliability, reproducibility, and interpretability of model outputs, which could ultimately reduce overall confidence in the results. As a result, model verification, validation, and uncertainty quantification (VVUQ) are essential for successful clinical translation of TDT technologies into RPT clinics.

This work addresses the need for rigorous VVUQ by introducing methods tailored to PBPK model development and implementation. We begin by describing the basic elements of PBPK modeling and parameter estimation. We then summarized the key methods used in the VVUQ pipeline. We also discuss additional VVUQ challenges in integrating PBPK-based dosimetry into the dynamic TDT pipeline and highlight areas for future work.

## 2. PBPK Modeling in RPTs

PBPK models for RPTs mathematically describe changes in the activity of a radiotheranostic agent over time within tissues (An overview of PBPK and PopPK Models: Applications to Radiopharmaceutical Therapies for Analysis and Personalization, Deni Hardiansyah et al. PET Clinics, 2026) (10, 11). PBPK modeling has an impressive track record as a versatile tool for predicting biodistribution, thereby supporting the selection and optimization of radiolabeled targeting agents (12, 13), and PK analysis in preclinical and exploratory clinical research settings. Using patient-specific inputs (14), particularly from PET or SPECT imaging and biofluid sampling (15), for parameter estimation, PBPK models can be used to optimize dosing regimen, including the injected activity, imaging time points (16), number of cycles and treatment intervals, to improve outcomes (17).

In addition, connecting radiobiology as a Pharmacodynamic modeling component with the PK structure of PBPK models enables the prediction of the exposure-response relationships of radiopharmaceuticals in target tissues.

This approach supports the simulation of treatment response over time. It links the effect of activity accumulation in tumors and normal tissues to tumor control probability and normal tissue complication probability, offering a practical tool for treatment planning and optimization (8, 18, 19). However, before deployment, any PBPK model must be thoroughly validated in the context of its intended use.

### *2.1 Framework and Parameterization of PBPK Models*

PBPK model in RPTs comprises three key components: (1) computational framework, (2) anatomical-physiological framework, and (3) radiopharmaceutical framework (physicochemical and PK properties of the radiopharmaceutical), as shown in Figure 1 (20). In addition, by specifying elements such as dosing regimen, imaging schedule, or sampling time points, the study design and treatment protocol help define which parameters need to be estimated, what level of model complexity is appropriate, and how the model should be calibrated or validated for the intended use (21, 22).

The computational framework includes the model code, execution environment, and control settings such as estimation algorithms and convergence criteria, which together define and solve the system of equations (10). The anatomical-physiological framework incorporates patient-dependent parameters that describe human physiology for the intended population and application. This consists of relevant tissue compartments interconnected by arterial and venous blood circulation, with tissue-specific volumes and blood flow rates. For example, the whole-body PBPK model for [$^{177}$Lu]Lu-DOTATATE consists of tumors, kidneys as the organ at risk and excreting organ, gastrointestinal track, bone, red marrow, brain, heart, liver, lungs, muscle, skin, spleen, prostate and adrenal glands compartments (23). The remaining body tissues are lumped into a single compartment to ensure that blood circulation is represented comprehensively in the model. Additionally, sub-compartments of some tissues, such as vascular, interstitial or intracellular spaces, may require additional tissue-specific properties, such as permeability surface area products and fractional volumes (24). The distribution of the radiopharmaceutical inside the body can be perfusion- (25) or permeability-limited (26), based on model assumptions and the intended application. Additional physiological parameters, such as plasma protein binding (27), hematocrit, and glomerular filtration rate, are also included to characterize radiopharmaceutical kinetics.

The radiopharmaceutical framework of a PBPK model describes the physicochemical properties of the radiolabeled agent, such as molecular size, charge, lipophilicity and chelation stability, which influence its absorption, distribution, metabolism, and elimination (ADME). Additional key considerations in this framework include the physical half-life of the conjugated radionuclide and the biodistribution, retention, or clearance of its radioactive decay products. The model also accounts for the binding kinetics of the targeting agents, including the competition between radiolabeled and unlabeled compounds for available binding sites (28). These properties are application-specific and depend on the radionuclide and targeting agent.

For any PBPK model, parameter values should be annotated to indicate whether they are derived from preclinical biodistribution studies, in vitro assays, pre-therapeutic imaging, or literature. Parameters estimated through fitting should be well-documented, including their mean and standard deviation (SD). Given sparse, heterogeneous biokinetic data in RPTs, various resampling and data-handling strategies have been used to enable model verification and validation, as discussed in the following sections.

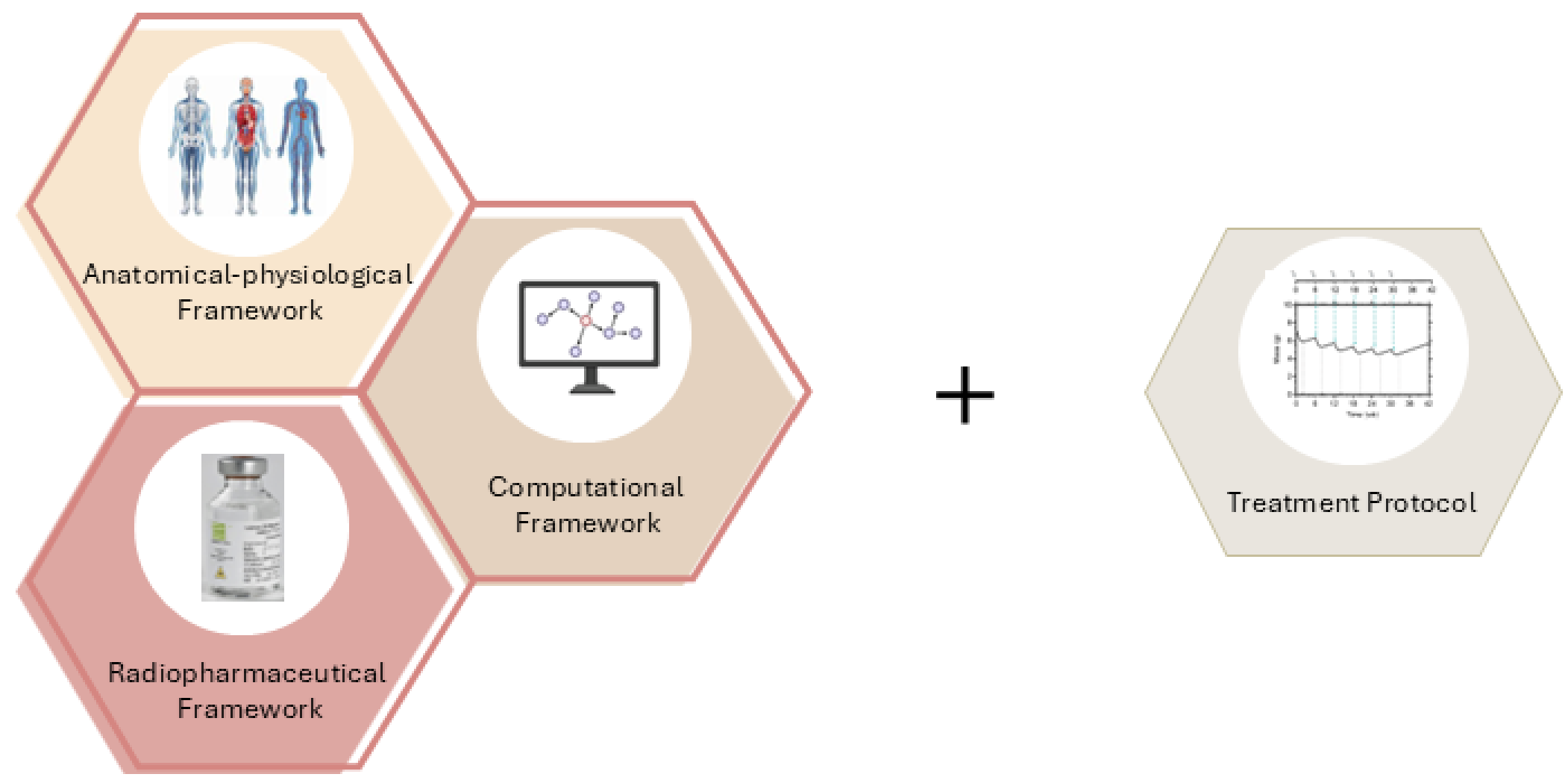

Figure 1: The building blocks of PBPK models in the context of RPTs. PBPK model development integrates three frameworks: the anatomical–physiological framework (patient-specific knowledge), the radiopharmaceutical framework (physicochemical and PK properties) and the computational framework (model code, execution environment and fitting algorithm) along with the additional treatment protocol framework.

### *2.2 Approaches to building PBPK models*

PBPK models for RPTs can be developed using bottom-up, top-down, or middle-out approaches based on data availability and modeling objectives (20). In the bottom-up approach, models are informed by *in vitro* or *in vivo* data to characterize the clinical ADME of the radiopharmaceutical. The bottom-up approach is typical for newly developed radiopharmaceuticals to inform early-phase trials. However, its reliability depends on input data quality and interspecies extrapolation, particularly for binding kinetics. These early-stage models often require additional refinement of patient-derived data from pre- and post-therapeutic imaging or sampling (26).

In contrast, the top-down approach estimates model parameters directly from patient data (29) and is typically used to describe uptake and clearance profiles. Interindividual variability of model parameters is captured through covariate selection, as shown in Siebinga et al.'s population approach to [$^{177}$Lu]Lu-PSMA-617, which used the covariate selection to evaluate the impact of different injected activities on salivary gland and tumor dosimetry (30). A key strength of top-down models is that they are well-suited for interpolating within the training data range; however, extrapolating remains challenging (20).

The middle-out approach integrates bottom-up mechanistic modeling with top-down data-driven calibration, making it well-suited for RPTs. By combining prior knowledge of physiological and drug characteristics, derived from measurements or literature, with patient-specific biokinetic data, it enables robust parameter optimization even in data-limited settings, a common condition in RPT, thereby supporting personalized treatment planning. For example, Hardiansyah et al. implemented a middle-out, population-based fitting of a whole-body PBPK model to estimate time-integrated activity coefficients (TIACs) and ADs, enabling accurate dosimetry from limited patient biokinetic data in peptide receptor radionuclide therapy (31).

## 3. VVUQ for PBPK Models in RPTs

Inaccuracy and imprecision in model predictions, such as TIAC, affect the reliability of PBPK models for dosimetry. While some sources of uncertainty (e.g., interpatient variability) are inherent, others arise from limitations in data, assumptions, or model structure and influence the confidence in model predictions. These can be reduced through improved data quality and iterative model development, supported by systematic VVUQ. Such steps are essential for reliable clinical decision-making. Various methods support VVUQ of PBPK models, each addressing specific components of model development and implementation as outlined in the proposed workflow in Table 1.

Verification confirms that the conceptual model (the anatomical-physiological framework, radiopharmaceutical framework of the PBPK model, and treatment protocol of the simulated dosing regimen) is correctly implemented in the computational framework. This step ensures that the model is appropriately implemented − in other words, that we are “building the model right” (32, 33). Validation evaluates whether the PBPK model reliably describes patient biokinetics for its intended use − that is, whether we are “building the right model” (32, 33). The final step is uncertainty quantification, in which the precision and accuracy of model predictions are evaluated to help balance therapeutic efficacy and toxicity and enable personalized RPT by accounting for biological variability.

Table 1: Workflow of model verification, validation and uncertainty quantification for PBPK models in RPTs

| VVUQ | Steps | Description |
| --- | --- | --- |
| I. Verification | 1. Model structure verification | Model structure is confirmed to reflect relevant physiology, physicochemical and PK properties of the radiopharmaceutical. |
| | 2. Implementation and performance verification | Model equations, parameters, unit consistency and simulations are checked. |

| | | | |
|---|---|---|---|
| | | 3. Goodness-of-fit assessment | Model predictions are assessed for their resemblance to biokinetic data for the intended use.<br>Goodness-of-fit tests include visual inspection of the fitted graphs, coefficient of variation of the fitted parameters, visual predictive checks and normalized prediction distribution errors. |
| II. | Internal Validation | 4. Model selection | The model that best describes patient biokinetic data is selected using Akaike Information Criterion, Bayesian Information Criterion, or F-test analyses. |
| | | 5. Sensitivity analysis | Pharmacokinetic parameters with the highest impact on model outputs are identified using global or local sensitivity analysis. |
| III. | External Validation | 6. *Prediction Evaluation* | Model generalizability is tested by evaluating its predictive performance beyond the dataset used for optimization. This is done using independent data, when available, or through data splitting and resampling techniques such as cross-validation |
| | | 7. Assignment of statistical distributions | Precision of model parameters and output are characterized by their mean, standard deviation, and associated confidence intervals. |
| | | 8. Sampling of the input space | Parameter uncertainty is propagated using sampling-based techniques. |
| IV. | Uncertainty quantification (Precision and accuracy analyses) | 9. Statistical characterization of outputs and accuracy assessment | The distribution of model outputs is analyzed using the standard statistical tools and compared with reference values using relative deviation, root-mean square error, or absolute deviation. |
| | | 10. Output variability evaluation | Confidence intervals and uncertainty quantification metrics provide insight into the reliability of model predictions. |

### ***3.1 Verification***

Verification begins with a conceptual check to ensure that the model structure reflects relevant anatomy, physiology, and pharmacokinetics, guided by literature, expert consensus, and alignment with the treatment protocol and intended objectives. Once confirmed, the next step is to ensure the correct implementation of the computational model. This involves reviewing mathematical equations, parameter definitions, and flow logic. The differential equations should maintain mass balance across all compartments, and all parameters must be expressed in consistent, physiologically appropriate units (33).

Visual inspection of the model behavior is an important aspect of verification. Plotting time-activity curves (TAC) for key organs and comparing them to expected, literature-based biokinetics, such as renal clearance or tumor uptake, can reveal inconsistencies, including abnormal, non-physiological accumulation patterns or clearance rates outside of reported ranges. Running the model under simplified conditions allows targeted evaluation of specific components; for example, disabling clearance pathways helps assess equilibrium under passive distribution. In addition, expert reviews of the model outputs can identify biologically implausible behavior that is not evident from the model structure. These layered verification strategies (Figure 2) ensure that the PBPK model is correctly developed and provide a robust foundation for subsequent validation and clinical applications in radiotheranostic settings.

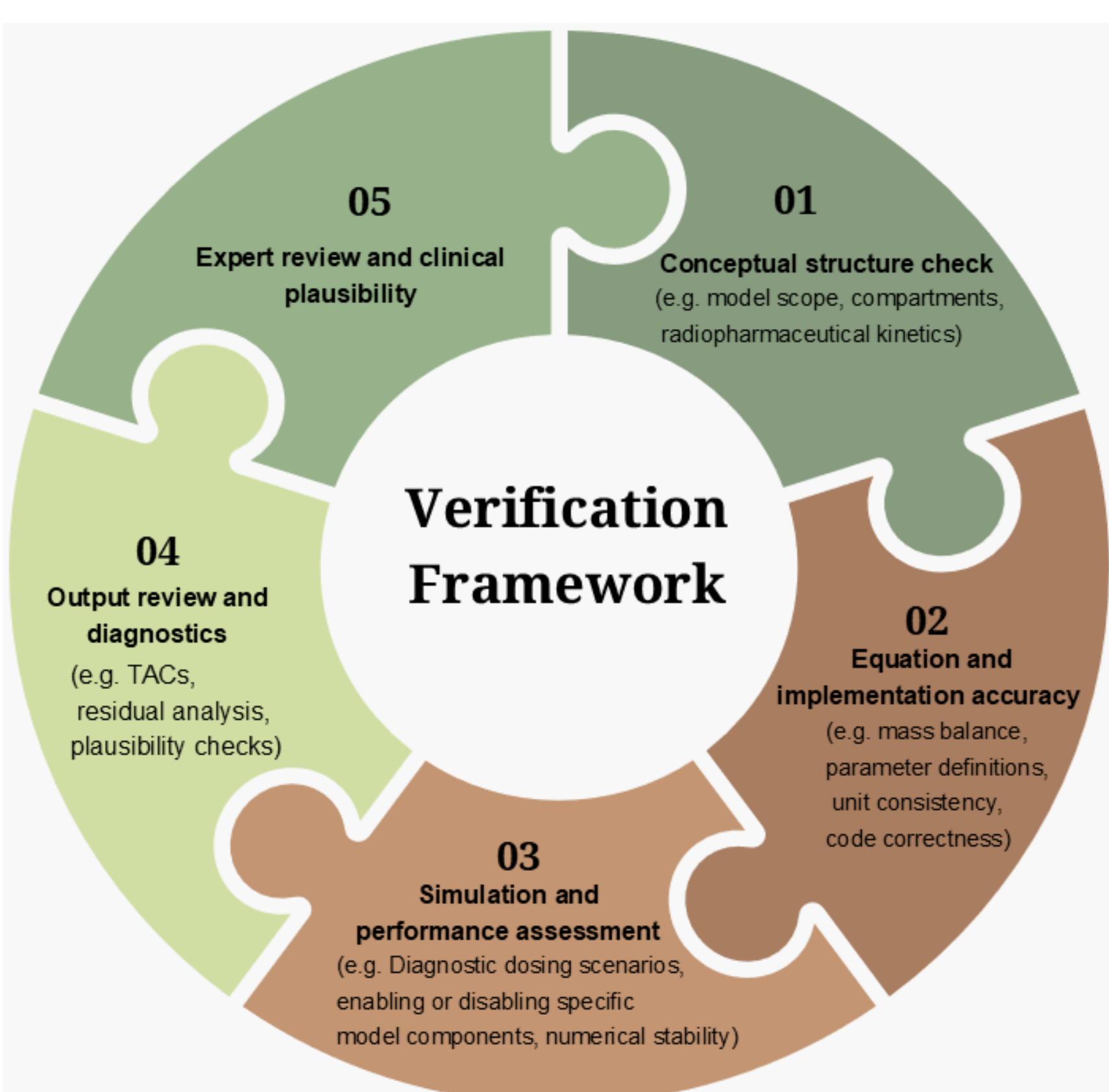

Figure 2: PBPK model verification framework. Recommended steps for “building the model right”.

### *3.2 Validation*

PBPK model development is an iterative process of prediction, evaluation, and refinement. Validation begins by testing whether the model can explain the internal training data that is used for model development (34). This internal validation is done by assessing the goodness-of-fit between model output and patient biokinetic data from blood, urine, or PET/SPECT imaging to ensure reliability for the intended use (10). Predictive performance is then externally validated by comparing simulated biokinetics with independent datasets not used in training (20, 35, 36). This stepwise approach ensures reliable predictions for personalized planning and broader population-level applications to assess generalizability and robustness.

### ***Internal Validation***

#### *3.2.1 Goodness-of-Fit Criteria (GOF)*

Evaluating the GOF is an essential step in PBPK model development, used to assess how well the model replicates the observed biokinetic data for the intended application (37). The residual differences between predicted values and measured data should be minimal, randomly distributed, and free of systematic bias. This suggests that the model adequately captures the PK and supports its use for further simulations or predictions. Several approaches are used to evaluate GOF, including graphical diagnostics and quantitative metrics to guide clinical interpretation and model refinement (Table 2) (37).

#### *3.2.2 Simulation-Based Evaluation*

Posterior predictive checks evaluate the agreement between PBPK model simulations and biokinetic data, assuming the model sufficiently captures the PK of the radiopharmaceuticals (38). Simulations are conducted using the original dosing regimen and sampling schedule to generate a distribution of predicted outputs, such as TIACs, across body tissues. Confidence intervals (CIs) from simulations are compared to patient data using metrics such as mean prediction error. Visual predictive check and normalized prediction distribution errors are commonly used to evaluate variability across patients (39, 40).

Table 2: Model validation: Graphs and metrics (10, 37, 41).

| GOF Criteria | Summary of the techniques | Equation/Plot |
| --- | --- | --- |
| Visual inspection of the fitted curves | Verify that the fitted curve passes through or near most biokinetic data points and displays no systemic trends. | |
| Observed vs. Predicted plots | Compare predicted values with observed data using scatter plots. Alignment along the identity line indicates good model accuracy. | Predicted<br>Observed |
| Residual vs. Predicted or Time Plots | Examine residuals plotted against predictions or time. A random scatter with no discernible pattern suggests unbiased residuals and a good fit. | 0.010<br>0.005<br>0.000<br>-0.005<br>-0.010<br>Residual<br>Predicted |
| Histograms of Residuals | Inspect the distribution of residuals. A bell-shaped histogram centered around zero supports the assumption of normally distributed and unbiased errors. | 40<br>30<br>20<br>10<br>0<br>Frequency<br>-80 -60 -40 -20 0 20 40 60 80 |
| Sum of Squares Error (SSE) | Calculate the total squared differences between observed ($\dot{y}$) and predicted (y) values. Lower SSE values indicate better overall fit to the data. | $SSE = \sum_{i=1}^{F} (y - \dot{y})^2$ |
| Mean Squared Error (MSE) | Measures the average squared difference between observed and predicted values. Lower MSE values indicate better model performance. | $MSE = \frac{SSE}{n - p}$<br>n: data points<br>p: estimated parameters |
| Coefficient of Determination ($R^2$) | Assesses how much of the variability in the observed data is explained by the model. Values near 1 suggest strong explanatory power. | $R^2 = 1 - \frac{SSE}{\sum_{i=1}^{F} (y - \bar{y})^2}$<br>$\bar{y}$: mean value of $y$ |
| Coefficients of Variation (CV) | Evaluates the precision of the estimated parameters.<br>CV<25%: precise<br>CV<50%: acceptable.<br>CV>50%: poor identifiability | $\text{CV}\ (q_i) = \frac{\sqrt{Var(q_i)}}{\bar{q_i}}$<br>$Var(q_i)$: variance of the estimated parameter $q_i$ |
| Correlation Matrix | Analyzes interparameter dependencies.<br>Low off-diagonal values (e.g., between -0.8 and 0.8) indicate low correlation and good parameter identifiability. | $corr(q_i, q_j) = \frac{cov(q_i, q_j)}{\sqrt{Var(q_i)\, Var(q_j)}}$<br>$corr(q_i, q_j)$ : correlation coefficient between the estimated parameters<br>$cov(q_i, q_j)$: covariance between parameters $q_i$ and $q_j$ |

*3.2.3 Model selection*

The selection of PBPK models balances biological plausibility with parameter identifiability, particularly in RPTs with sparse or heterogeneous data. The principle of parsimony recommends models no more complex than what the data can support. While overparameterized models may closely fit biokinetic data, they risk capturing noise rather than true physiological processes, reducing generalizability. Conversely, oversimplified models may miss key PK, such as tissue retention or nonlinear binding (37).

To address this, the Akaike Information Criterion (AIC) and its small-sample correction (AICc) are widely used for model selection (42). For example, AICc supported selecting bivalent over monovalent binding to better describe $^{111}$In-labeled anti-CD66 biodistribution (24). The Akaike weight of model ($i$) ($w_{AICci}$) that represents the probability that model $i$ is the best among those considered is calculated as follows:

$$AICc = -2\, ln(P) + 2K + \frac{2K(K+1)}{N-K-1} \quad (1)$$

$$\Delta_i = AIC_{ci} - AIC_{cmin} \quad (2)$$

$$w_{AICci} = \frac{e^{-\frac{\Delta_i}{2}}}{\sum_{i=1}^{F} e^{-\frac{\Delta_i}{2}}} \quad (3)$$

where P is the estimated objective function minimized during fitting, $K$ is the number of fitted parameters, $N$ is the number of data points, $AICc_{min}$ is the lowest $AICc$ among all models, $\Delta_i$ is the difference between the $AICc_i$ of model $i$ and $AICc_{min}$, F is the total number of candidate models.

Although the Bayesian Information Criterion (BIC) is more stringent in penalizing model complexity, its use in radiopharmaceutical applications is limited by small sample sizes and high intersubject variability (37). In such cases, the F-test offers a practical alternative for comparing nested models by assessing whether improved fit justifies added complexity (42). It is applied stepwise, testing models with increasing parameters, and selecting the best-fit model based on statistical significance (e.g., $p < 0.05$).

*3.2.4 Sensitivity Analysis (SA)*

SA quantifies how the uncertainty in input parameters affects PBPK model outputs (43, 44). Parameters that contribute significantly to output variability can be prioritized for targeted measurements or estimations to improve individualized dose calculations. Conversely, low-impact parameters may be fixed to simplify the model and reduce computational burden.

SA methods are generally categorized into local (one-at-a-time, OAT) or global (GSA) (44). OAT methods vary one parameter at a time while the others remain fixed. For example, Abdollahi et al. used OAT in a PBPK model of [$^{177}$Lu]Lu-PSMA to evaluate how receptor density, ligand amount, and internalization rate affect tumor and kidney ADs (19). Their results demonstrated that tumor uptake was particularly sensitive to PSMA expression and ligand amount. Although simple, OAT may overlook nonlinearities and parameter interactions common in PBPK models (43).

GSA quantifies the relative influence of input parameters on model outputs by systematically varying all parameters over their predefined probability distributions, typically through Monte Carlo or Latin hypercube sampling, and analyzing the resulting variance or distribution of the outputs (45). GSA begins by specifying the uncertain parameters along with their probability distributions and any dependencies, generating a large ensemble of parameter sets that spans the joint input space, and propagating each set through the model to obtain the corresponding output ensemble. Parameter influence is then quantified from these simulations using global metrics such as variance-based Sobol first-order and total-effect indices, moment-independent measures, or screening approaches, thereby attributing output variability to each input over its entire range. Commonly applied GSA methods include the Sobol and Fourier Amplitude Sensitivity Test (FAST) approaches, the Morris screening method, distribution-based techniques such as multi-parametric sensitivity analysis, and extensions that account for correlated or dependent inputs, including the Kucherenko method (44, 45). Hardiansyah et al. applied GSA using the extended FAST (eFAST) in a [$^{177}$Lu]Lu-PSMA PBPK model (46). The eFAST method was chosen because it requires fewer model evaluations and imposes a lower computational burden compared with the Sobol and original FAST methods. Their analysis showed renal dose variability was most influenced by kidney blood flow and receptor density, while tumor dose was most sensitive to receptor density. This highlights how GSA improves interpretability and clinical relevance by identifying parameters as key contributors to absorbed dose variability in [$^{177}$Lu]Lu-PSMA therapy, supporting better data collection, dose estimation, and personalized treatment decisions.

***External Validation***

3.2.5 Prediction Evaluation

Data splitting and resampling techniques can be used to evaluate the confidence in PBPK model performance. Data splitting divides biokinetic datasets into calibration and evaluation subsets to assess generalizability. Resampling methods, such as bootstrapping, cross-validation, or the jackknife approach, facilitate uncertainty quantification in parameter estimates (38). For example, Budiansah et al. applied a leave-one-out jackknife resampling in a PBPK study of [$^{111}$In]In-DOTA-TATE as a surrogate for [$^{90}$Y]Y-DOTA-TATE to evaluate the feasibility of using single-time-point (STP) imaging measurement for dose estimation (42). These approaches work even with limited data, reinforcing the role of PBPK models in model-informed radiopharmaceutical therapy.

### *3.3 Uncertainty Quantification (UQ)*

Systematic UQ for PBPK models is essential for reliable, patient-specific dosimetry in RPTs. By characterizing confidence in TIA estimates, UQ supports treatment strategies that optimize efficacy while limiting toxicity within clinical thresholds (47). Given the inherent variability in physiological parameters and biokinetic data, UQ provides a framework for quantifying CIs around model predictions. This is especially important in PBPK models, where structural or parameter uncertainties can compromise predicted outcomes (48, 49). UQ enhances the robustness of treatment planning and supports adaptive strategies that accommodate inter- and intrapatient variability over time.

The process of UQ begins with assigning appropriate statistical distributions to key PK parameters, defined by their mean, SD, and CIs based on literature, fitting, or experimental data (50). Parameter variability propagates during sampling methods, such as Latin Hypercube Sampling or Monte Carlo simulations (49). For each sample set, model simulations generate outputs of interest with their variance, such as TIAC and prediction precision are statistically characterized using the standard statistical tools and CIs (49). Additional metrics evaluate prediction accuracy relative to reference values. Relative deviation (RD) quantifies the bias, and the SD of the RD reflects variability across patients or time points (37, 49). The root-mean square error (RMSE) combines the mean and SD of the RD. Mean absolute percentage error (MAPE) captures the average deviation and is less sensitive to outliers than RMSE, which can be determined as shown in the following equations (31, 51):

$$RD_i = \frac{Output_i - Output_r}{Output_r} \tag{4}$$

$$RMSE_i = \sqrt{(SD\ of\ RD_i)^2 - (Mean\ of\ RD_i)^2} \tag{5}$$

$$MAPE = \frac{1}{n}\sum_{i=1}^{n}|RD_i| \tag{6}$$

where $Output_i$ is the result of model simulations, while the corresponding reference value is $Output_r$. $n$ is the number of patients.

Finally, prediction variability is assessed against clinical thresholds (e.g., organ toxicity limits) to inform treatment planning.

Budiansah et al. evaluated the accuracy and precision of STP dosimetry using nonlinear mixed-effects modeling (NLMEM) implemented in a PBPK model for [ $^{111}$In]In-DOTA-TATE. ADs from STP data were compared to multiple time-point references using RD and RMSE, while precision was assessed via relative standard errors. STP dosimetry at 48 hours post-injection showed the highest accuracy of the predicting AD, though with lower precision than multiple time point dosimetry (51).

## 4. Extending PBPK modeling to dynamic TDTs: emerging VVUQ challenges

PBPK models in RPTs provide a mechanistic approach for estimating the radiopharmaceutical biodistribution based on fixed and optimized inputs. However, PBPK models are only one part of the larger CNO pipeline for building TDTs. The TDTs framework includes three core components (Figure 3): the physical object (actual patient), the virtual representation (patient *in silico*), and a bi-directional feedback loop that enables real-time data acquisition and human-in-the-loop clinical decision making across physical and virtual domains (52). This exchange keeps the model aligned with the patient's current physiological state using repeated imaging, laboratory values, and clinical observations (53).

TDTs integrate radiobiological models and PK parameters with PBPK-derived tissue dosimetry, including small-scale dosimetry that considers the dynamic and spatial heterogeneity of tumors and normal tissues. This heterogeneity can be incorporated into dose-volume histogram (DVH) analyses to evaluate the interpatient variability of biological effective dose (BED). To enable such analysis, the TDTs should combine the PBPK framework with additional modular components and support the dynamic integration of updated patient-specific data throughout treatment (15). For example, renal clearance may be estimated using serum creatinine and glomerular filtration rate, while hematologic status can be evaluated via neutrophil and platelet counts. Biomarkers, such as tumor volume or prostate-specific antigen (PSA) levels in prostate cancer, may be integrated to capture changes in treatment response, such as tumor volumes. These continuous inputs allow TDTs to remain aligned with evolving PK and therapeutic response over time (54).

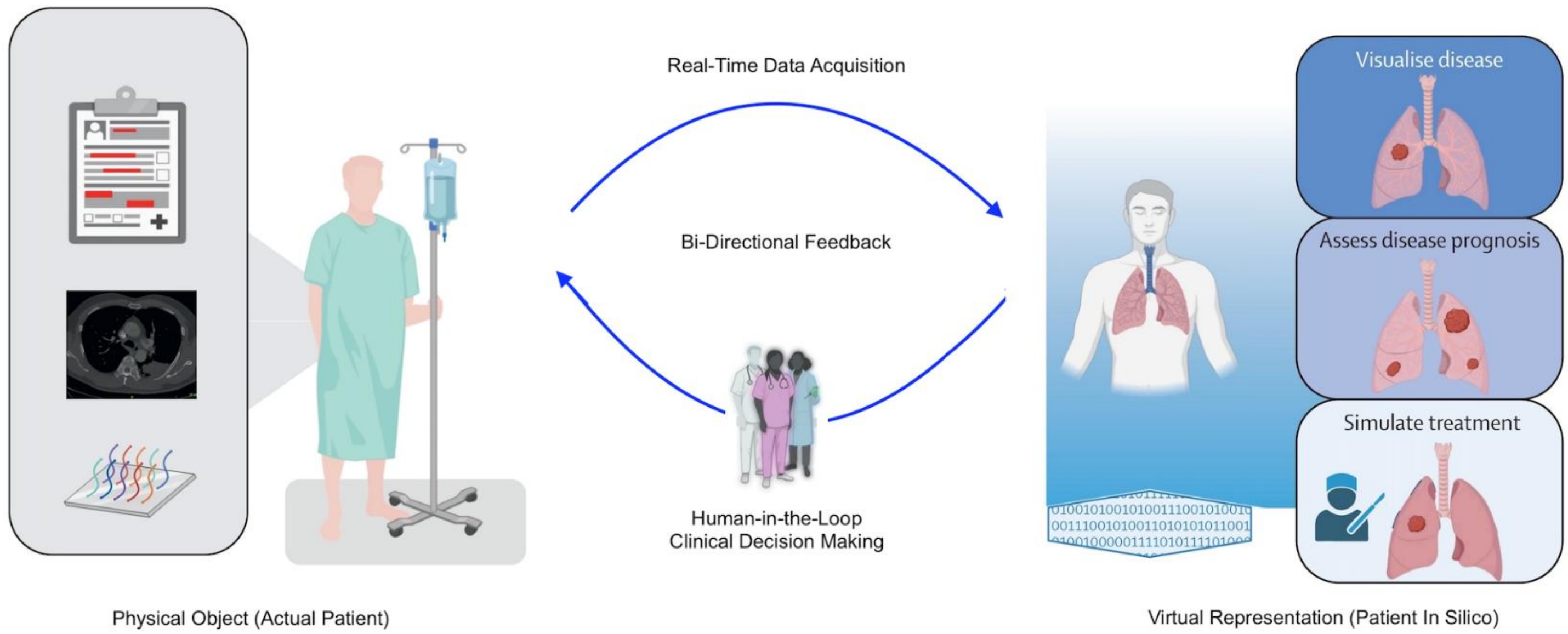

Figure 3: Basic components of biomedical digital twins, including TDTs based on the CNO pipeline. VVUQ must be integrated to ensure the reliability of TDTs predictions for clinical decision making. Diagram adapted from (55).

Information flowing from the TDT to the clinic includes patient-specific predictions of AD to tissues, assessments of toxicity, and evaluations of alternative treatment regimens. These outputs support prospective clinical decision-making, including the selection of administered activity, adjustment of treatment intervals, and protocol adaptation based on response (Multiscale Computational Radiobiology for Precision Radiopharmaceutical Therapies Tahir Yusufaly et al. PET Clinics, 2026). Integrating TDTs into clinical workflows enables a shift from empirical, population-based dosing to model-informed approaches that reflect the unique biological characteristics of each patient (52, 53).

TDTs and biomedical DTs present distinct challenges to VVUQ due to their dynamic and continuously updated virtual representation of the human body. These processes of VVUQ must be iterative, data-responsive, and tightly coupled to the temporal evolution of the patient-specific system (36). Whereas static models are constructed and verified using fixed datasets under controlled conditions, TDTs depend on the ongoing integration of quantitative imaging data and patient-specific physiological inputs (56). This continuous inflow introduces variability not only in parameters of interest but also in imaging protocols, scanner calibration, TAC resolution, and the fidelity of quantitative SPECT or PET measurements. As a result, verification must ensure not only computational correctness but also the compatibility with incoming clinical data, which may be affected by hardware differences, acquisition timing, or tracer quantification noise (36). TDTs also require continuous validation to reflect evolving radiopharmaceutical PKs due to lesion progression, normal organ toxicity, or treatment response. For example, internal validation is used to confirm that the first-cycle biokinetic data are reproduced by TDT simulations using the evaluation criteria in Table 2. External validation assesses predictive accuracy by comparing simulated second-cycle kinetics and organ/lesion doses against observed values, verifying that they fall within prediction intervals. Results from external validation are then used to iteratively refine the model for subsequent cycles, maintaining a continuous and dynamic update loop of TDT across treatment. Maintaining predictive accuracy and consistent quantification of associated uncertainty for TDTs in the presence of sparse, noisy, or delayed imaging inputs is challenging, but essential. This dynamic context challenges generalizability, especially in sparse and heterogeneous patient-specific imaging data.

## 5. Summary

Model-informed personalized treatment planning is essential for optimizing therapeutic outcomes, especially following the FDA's approval of [$^{177}$Lu]Lu-PSMA for earlier-stage treatment. CNO modeling serves as a mechanistic framework predicting ADs to tissues and treatment responses. Confidence in CNO models relies on rigorous VVUQ: verification ensures proper model implementation; validation confirms agreement with measured biokinetics; and uncertainty quantification captures variability in predictions.

Despite their potential, PBPK models for RPTs remain underdeveloped from a regulatory perspective. While PBPK is widely accepted in the pharmaceutical industry and supported by the FDA and European Medicines Agency (EMA) guidance (57, 58), comparable standards for radiopharmaceuticals have yet to be defined.

Establishing clear, fit-for-purpose VVUQ frameworks will strengthen confidence in PBPK model applications, support regulatory approval in RPTs, and promote consistent integration into clinical workflows. Aligning these efforts with ongoing precision medicine initiatives and engaging professional organizations such as the Society of Nuclear Medicine and Molecular Imaging (SNMMI), European Association of Nuclear Medicine (EANM), and International Atomic Energy Agency (IAEA), will be critical for advancing the translation of PBPK models in RPTs from research to clinical practice.

## References

1. Herrmann K, M.D., Schwaiger M, M.D., Lewis JS, PhD., Solomon SB, M.D., McNeil BJ, M.D., Baumann M, M.D., et al. Radiotheranostics: a roadmap for future development. The lancet oncology. 2020;21(3):e146–56.

2. Weber WA, Barthel H, Bengel F, Eiber M, Herrmann K, Schäfers M. What Is Theranostics? Journal of Nuclear Medicine. 2023;64(5):669–70.

3. Stokkel MPM, Gotthardt M, Herrmann K, Gnanasegaran G. Theranostics in Perspective: White Paper. Journal of Nuclear Medicine. 2025:jnumed.125.269776.

4. Fallah J, Agrawal S, Gittleman H, Fiero MH, Subramaniam S, John C, et al. FDA Approval Summary: Lutetium Lu 177 Vipivotide Tetraxetan for Patients with Metastatic Castration-Resistant Prostate Cancer. Clinical cancer research. 2023;29(9):1651–7.

5. Sartor O, de Bono J, Chi KN, Fizazi K, Herrmann K, Rahbar K, et al. Lutetium-177–PSMA-617 for Metastatic Castration-Resistant Prostate Cancer. N Engl J Med. 2021;385(12):1091–103.

6. Strigari L, Schwarz J, Bradshaw T, Brosch-Lenz J, Currie G, El-Fakhri G, et al. Computational Nuclear Oncology Toward Precision Radiopharmaceutical Therapies: Ethical, Regulatory, and Socioeconomic Dimensions of Theranostic Digital Twins. Journal of Nuclear Medicine. 2025;66(5):748–56.

7. Jiménez-Franco LD, Kletting P, Beer AJ, Glatting G. Treatment planning algorithm for peptide receptor radionuclide therapy considering multiple tumor lesions and organs at risk. Medical physics (Lancaster). 2018;45(8):3516–23.

8. Kletting P, Thieme A, Eberhardt N, Rinscheid A, D'Alessandria C, Allmann J, et al. Modeling and Predicting Tumor Response in Radioligand Therapy. Journal of Nuclear Medicine. 2019;60(1):65–70.

9. Zaid NRR, Bastiaannet R, Hobbs R, Sgouros G. Mathematic Modeling of Tumor Growth During [ 177 Lu]Lu-PSMA Therapy: Insights into Treatment Optimization. Journal of Nuclear Medicine. 2025;66(1):84–90.

10. Ivashchenko OV, O'Doherty J, Hardiansyah D, Cremonesi M, Tran-Gia J, Hippeläinen E, et al. Time-Activity data fitting in molecular Radiotherapy: Methodology and pitfalls. Physica medica. 2024;117:103192.

11. Sun Z, Zhao N, Zhao X, Wang Z, Liu Z, Cui Y. Application of physiologically based pharmacokinetic modeling of novel drugs approved by the U.S. food and drug administration. European journal of pharmaceutical sciences. 2024;200:106838.

12. Zaid NRR, Kletting P, Beer AJ, Stallons TAR, Torgue JJ, Glatting G. Mathematical Modeling of In Vivo Alpha Particle Generators and Chelator Stability. Cancer Biother Radiopharm. 2023;38(8):528–35.

13. Begum NJ, Glatting G, Wester H, Eiber M, Beer AJ, Kletting P. The effect of ligand amount, affinity and internalization on PSMA-targeted imaging and therapy: A simulation study using a PBPK model. Sci Rep. 2019;9(1):20041–8.

14. Rahmim A, Brosch-Lenz J, Fele-Paranj A, Yousefirizi F, Soltani M, Uribe C, et al. Theranostic digital twins for personalized radiopharmaceutical therapies: Reimagining theranostics via computational nuclear oncology. Frontiers in oncology. 2022;12:1062592.

15. Abdollahi H, Yousefirizi F, Shiri I, Brosch-Lenz J, Mollaheydar E, Fele-Paranj A, et al. Theranostic digital twins: Concept, framework and roadmap towards personalized radiopharmaceutical therapies. Theranostics. 2024;14(9):3404–22.

16. Rinscheid A, Kletting P, Eiber M, Beer AJ, Glatting G. Influence of sampling schedules on [177Lu]Lu-PSMA dosimetry. EJNMMI Phys. 2020;7(1):41.

17. Siebinga H, de Wit-van der Veen BJ, Stokkel MDM, Huitema ADR, Hendrikx JJMA. Current use and future potential of (physiologically based) pharmacokinetic modelling of radiopharmaceuticals: a review. Theranostics. 2022;12(18):7804–20.

18. DERENDORF H, MEIBOHM B. Modeling of pharmacokinetic/pharmacodynamic (PK/PD) relationships : Concepts and perspectives. Pharm Res. 1999;16(2):176–85.

19. Abdollahi H, Fele-Paranj A, Rahmim A. Model-Informed Radiopharmaceutical Therapy Optimization: A Study on the Impact of PBPK Model Parameters on Physical, Biological, and Statistical Measures in 177Lu-PSMA Therapy. Cancers. 2024;16(18):3120.

20. Shebley M, Sandhu P, Emami Riedmaier A, Jamei M, Narayanan R, Patel A, et al. Physiologically Based Pharmacokinetic Model Qualification and Reporting Procedures for Regulatory Submissions: A Consortium Perspective. Clin Pharmacol Ther. 2018;104(1):88–110.

21. Abouir K, Samer CF, Gloor Y, Desmeules JA, Daali Y. Reviewing Data Integrated for PBPK Model Development to Predict Metabolic Drug-Drug Interactions: Shifting Perspectives and Emerging Trends. Frontiers in pharmacology. 2021;12:708299.

22. Kuepfer L, Niederalt C, Wendl T, Schlender J, Willmann S, Lippert J, et al. Applied Concepts in PBPK Modeling: How to Build a PBPK/PD Model. CPT: pharmacometrics and systems pharmacology. 2016;5(10):516–31.

23. Vasić V, Gustafsson J, Nowshahr EY, Stenvall A, Beer AJ, Gleisner KS, et al. A PBPK model for PRRT with [177Lu]Lu-DOTA-TATE: Comparison of model implementations in SAAM II and MATLAB/SimBiology. Physica medica. 2024;119:103299.

24. Kletting P, Kull T, Bunjes D, Mahren B, Luster M, Reske SN, et al. Radioimmunotherapy with Anti-CD66 Antibody: Improving the Biodistribution Using a Physiologically Based Pharmacokinetic Model. Journal of Nuclear Medicine. 2010;51(3):484–91.

25. Kletting P, Maaß C, Reske S, Beer AJ, Glatting G. Physiologically Based Pharmacokinetic Modeling Is Essential in 90Y-Labeled Anti-CD66 Radioimmunotherapy. PloS one. 2015;10(5):e0127934.

26. Kletting P, Schuchardt C, Kulkarni HR, Shahinfar M, Singh A, Glatting G, et al. Investigating the Effect of Ligand Amount and Injected Therapeutic Activity: A Simulation Study for 177Lu-Labeled PSMA-Targeting Peptides. PloS one. 2016;11(9):e0162303.

27. Fele-Paranj A, Saboury B, Uribe C, Rahmim A. Correction: Physiologically based radiopharmacokinetic (PBRPK) modeling to simulate and analyze radiopharmaceutical therapies: studies of non-linearities, multi-bolus injections, and albumin binding. EJNMMI radiopharm chem. 2024;9(1):22.

28. Zaid NRR, Kletting P, Winter G, Prasad V, Beer AJ, Glatting G. A Physiologically Based Pharmacokinetic Model for In Vivo Alpha Particle Generators Targeting Neuroendocrine Tumors in Mice. Pharmaceutics. 2021;13(12):2132.

29. Siebinga H, de Wit-van der Veen BJ, Beijnen JH, Stokkel MPM, Dorlo TPC, Huitema ADR, et al. A physiologically based pharmacokinetic (PBPK) model to describe organ distribution of 68Ga-DOTATATE in patients without neuroendocrine tumors. EJNMMI Res. 2021;11(1):73.

30. Siebinga H, Privé BM, Peters SMB, Nagarajah J, Dorlo TPC, Huitema ADR, et al. Population pharmacokinetic dosimetry model using imaging data to assess variability in pharmacokinetics of 177Lu-PSMA-617 in prostate cancer patients. CPT: pharmacometrics and systems pharmacology. 2023;12(8):1060–71.

31. Hardiansyah D, Riana A, Beer AJ, Glatting G. Single-time-point estimation of absorbed doses in PRRT using a non-linear mixed-effects model. Zeitschrift für medizinische Physik. 2023;33(1):70–81.

32. Sargent RG. Verification and validation of simulation models. IEEE; 2010.

33. Robinson S, Healy KJ, Andradóttir S, Nelson BL, Withers DH. Simulation model verification and validation: increasing the users' confidence. Washington, DC, USA: IEEE Computer Society; 1997.

34. Aldridge BB, Burke JM, Lauffenburger DA, Sorger PK. Physicochemical modelling of cell signalling pathways. Nat Cell Biol. 2006;8(11):1195–203.

35. Peters SA, Dolgos H. Requirements to Establishing Confidence in Physiologically Based Pharmacokinetic (PBPK) Models and Overcoming Some of the Challenges to Meeting Them. Clin Pharmacokinet. 2019;58(11):1355–71.

36. Frechen S, Rostami-Hodjegan A. Quality Assurance of PBPK Modeling Platforms and Guidance on Building, Evaluating, Verifying and Applying PBPK Models Prudently under the Umbrella of Qualification: Why, When, What, How and By Whom? Pharm Res. 2022;39(8):1733–48.

37. Bonate PL. Pharmacokinetic-Pharmacodynamic Modeling and Simulation. 2nd ed. Boston, MA: Springer Science + Business Media; 2011.

38. Huang S, Ding Q, Yang N, Sun Z, Cheng Q, Liu W, et al. External evaluation of published population pharmacokinetic models of posaconazole. Frontiers in pharmacology. 2022;13:1005348.

39. Sherwin CMT, Kiang TKL, Spigarelli MG, Ensom MHH. Fundamentals of Population Pharmacokinetic Modelling: Validation Methods. Clin Pharmacokinet. 2012;51(9):573–90.

40. Nguyen THT, Mouksassi M, Holford N, Al-Huniti N, Freedman I, Hooker AC, et al. Model Evaluation of Continuous Data Pharmacometric Models: Metrics and Graphics. CPT: pharmacometrics and systems pharmacology. 2017;6(2):87–109.

41. Kletting P, Schimmel S, Kestler HA, Hänscheid H, Luster M, Fernández M, et al. Molecular radiotherapy: The NUKFIT software for calculating the time-integrated activity coefficient. Medical physics (Lancaster). 2013;40(10):102504,n/a.

42. Glatting G, Kletting P, Reske SN, Hohl K, Ring C. Choosing the optimal fit function: Comparison of the Akaike information criterion and the F-test. Medical physics (Lancaster). 2007;34(11):4285–92.

43. McNally K, Cotton R, Loizou GD. A Workflow for Global Sensitivity Analysis of PBPK Models. Frontiers in pharmacology. 2011;2:31.

44. Qian G, Mahdi A. Sensitivity analysis methods in the biomedical sciences. Math Biosci. 2020;323:108306.

45. Zi Z, Cho K, Sung M, Xia X, Zheng J, Sun Z. In silico identification of the key components and steps in IFN-γ induced JAK-STAT signaling pathway. FEBS Lett. 2005;579(5):1101–8.

46. Hardiansyah D, Kletting P, Begum NJ, Eiber M, Beer AJ, Pawiro SA, et al. Important pharmacokinetic parameters for individualization of 177Lu-PSMA therapy: A global sensitivity analysis for a physiologically-based pharmacokinetic model. Medical physics (Lancaster). 2021;48(2):556–68.

47. Stabin MG. Uncertainties in Internal Dose Calculations for Radiopharmaceuticals. Journal of Nuclear Medicine. 2008;49(5):853–60.

48. Barton HA, Chiu WA, Woodrow Setzer R, Andersen ME, Bailer AJ, Bois FY, et al. Characterizing Uncertainty and Variability in Physiologically Based Pharmacokinetic Models: State of the Science and Needs for Research and Implementation. Toxicological Sciences. 2007;99(2):395–402.

49. Spielmann V, Li WB, Zankl M, Oeh U, Hoeschen C. Uncertainty Quantification in Internal Dose Calculations for Seven Selected Radiopharmaceuticals. Journal of Nuclear Medicine. 2016;57(1):122–8.

50. Gear JI, Cox MG, Gustafsson J, Gleisner KS, Murray I, Glatting G, et al. EANM practical guidance on uncertainty analysis for molecular radiotherapy absorbed dose calculations. Eur J Nucl Med Mol Imaging. 2018;45(13):2456–74.

51. Budiansah I, Hardiansyah D, Riana A, Pawiro SA, Beer AJ, Glatting G. Accuracy and precision analyses of single-time-point dosimetry utilising physiologically-based pharmacokinetic modelling and non-linear mixed-effects modelling. EJNMMI Phys. 2025;12(1):26–15.

52. Katsoulakis E, Wang Q, Wu H, Shahriyari L, Fletcher R, Liu J, et al. Digital twins for health: a scoping review. npj Digit Med. 2024;7(1):77.

53. Sel K, Hawkins-Daarud A, Chaudhuri A, Osman D, Bahai A, Paydarfar D, et al. Survey and perspective on verification, validation, and uncertainty quantification of digital twins for precision medicine. npj Digit Med. 2025;8(1):40–12.

54. Bruynseels K, Santoni de Sio F, van den Hoven J. Digital Twins in Health Care: Ethical Implications of an Emerging Engineering Paradigm. Frontiers in genetics. 2018;9:31.

55. Sadée C, MRes, Testa S, M.D., Barba T,M.D.PhD., Hartmann K,M.D.PhD., Schuessler M,M.S.M.P.P., Thieme A,Dr med, et al. Medical digital twins: enabling precision medicine and medical artificial intelligence. The Lancet.Digital health. 2025:100864.

56. Olsson Gisleskog P, Hermann D, Hammarlund-Udenaes M, Karlsson MO. Validation of a population pharmacokinetic/pharmacodynamic model for 5α-reductase inhibitors. European journal of pharmaceutical sciences. 1999;8(4):291–9.

57. Zhao P. Report from the EMA workshop on qualification and reporting of physiologically based pharmacokinetic (PBPK) modeling and simulation. CPT: pharmacometrics and systems pharmacology. 2017;6(2):71–2.

58. Physiologically Based Pharmacokinetic Analyses - Format and Content Guidance for Industry2018.